\documentclass[aps,prd,superscriptaddress]{revtex4}
\usepackage{graphicx}
\usepackage{amssymb}
\usepackage{amsmath}
\usepackage{psfrag}

\newcommand{\ctg}{\mbox{cot}}

\begin{document}

\title{Off-diagonal correlations of lattice impenetrable bosons in
  one dimension}

\author{D.M.~Gangardt}
\email[e-mail: ]{gangardt@lptms.u-psud.fr}
\affiliation{Laboratoire de Physique Th\'eorique et Mod\`eles
  Statistiques, Universit\'e Paris Sud, 91405 Orsay Cedex, France}
\author{G.V.~Shlyapnikov}
\affiliation{Laboratoire de Physique Th\'eorique et Mod\`eles
  Statistiques, Universit\'e Paris Sud, 91405 Orsay Cedex, France}
\affiliation{Van der Waals - Zeeman Institute, University of Amsterdam, 
  Valckenierstraat 65/67, 1018 XE Amsterdam, The Netherlands}

\date{\today}

\begin{abstract}
We consider off-diagonal correlation functions of impenetrable bosons
on a lattice. By using the Jordan-Wigner transformation the one-body
density matrix is represented as (Toeplitz) determinant of a matrix of
fermionic Green functions. Using the replica method we calculate exactly
the full long-range  asymptotic behaviour of the one-body density matrix. 
We discuss how subleading oscillating terms, originating from short-range
correlations give rise to interesting features in the  momentum
distribution.   
\end{abstract}

\maketitle

\section{Introduction}
\label{sec:intro}

Correlation properties of one-dimensional bosons with short-range interactions
is a long standing problem of statistical physics of integrable systems
\cite{KorepinBook}. Despite the exact integrability, the correlation functions
are hard to calculate due to a complicated form of the Bethe Ansatz
wavefunctions and corresponding expressions for the matrix elements.  Recently
there has been a renewal of interest in these systems due to their relevance
for current experiments with cold atoms.  Experimental progress made it
possible to create strongly correlated bosonic states in the gas phase
\cite{Laburthe2004,Kinoshita2004} and in an optical lattice
\cite{Paredes2004,Koehl2004,Fertig2005}, and approach the Girardeau-Tonks
regime of impenetrable bosons. The physics of bosons in this regime is similar
in many respects to that of free fermions \cite{Girardeau60-65}, since the
strong repulsive interactions effectively play a role of the Pauli principle.
There is, however, an important difference due to bosonic quantum statistics
of particles. The statistics reveals itself in off-diagonal correlations of
bosons, such as the one-body density matrix or momentum distribution. The
calculation of these observables in the Girardeau-Tonks limit from the first
priciples represents an important question of statistical physics.

Impenetrable bosons in one dimension is an exactly soluble model both in the
continuum case and on a lattice. In the latter case it is equivalent to the
quantum XX spin chain: the two states of spin 1/2 on a given site correspond
to the presence/absence of a boson.  The correspondence with fermions was put
forward in the early work by Lieb, Schultz and Mattis \cite{Lieb1961}, where
this system was mapped onto a system of free fermions on a lattice by the
Jordan-Wigner transformation. Using this transformation it was possible to
represent transverse spin correlation functions (corresponding to the
off-diagonal correlations of bosonic operators) in the form of determinant of a
matrix with a size given by the correlation distance.  For a long time only an
approximate behaviour of these objects was available.  Schultz
\cite{Schultz1963} has shown that at zero temperature  correlation functions 
for large distances undergo a power law-decay. The explicit form of this power-law
decay has been found by Efetov and Larkin \cite{Efetov1975} and
has been  recovered later by Vaidya and Tracy \cite{Vaidya1978} 
using a method of mapping the quantum one-dimensional XX chain 
onto the two-dimensional classical Ising model. This
technique has been succesfully employed in the works of McCoy, Barouch and
Abraham \cite{McCoy1968-71} for calculating finite-temperature correlations.
Recent numerical studies \cite{Rigol2004} of impenetrable bosons on a lattice 
in the presence of a confining potential have shown a high degree 
of universality of the power law decay found in the uniform case.

In this paper we reconsider the problem of calculating off-diagonal
correlation functions of impenetrable bosons on a lattice. We start with the
Jordan-Wigner transformation and represent the one-body density matrix as a
determinant of a matrix given by fermionic Green functions. Next, we rewrite
the determinant in terms of an average over the ensemble of unitary random
matrices and calculate its asymptotic behaviour by employing the replica
method. Since its first use to derive the pair distribution function for
random matrices \cite{KamenevMezard1999}, the replica method has been
successfully applied to strongly correlated systems such as
Calogero-Sutherland models \cite{GangardtKamenev2001,Astrakharchik2006} and
impenetrable bosons in the continuum \cite{Gangardt2004}. The essence of the
replica method in this context is exactly the same as in the theory of
disordered systems \cite{Edwards1975}.  It consists of modifying the
quantities being averaged so that they depend on a parameter $n$ which is
assumed to be integer.Once the average is performed, the result is recovered
by a suitable analytical continuation to non-integer values of $n$.  The main
advantage of this method is that it leads to the asymptotic non-perturbative
result for the bosonic one-body density matrix including the
leading power-law term plus oscillating contributions reflecting the
short-range physics of the problem and lattice effects. This result allows one
to determine the main features of the momentum distribution of impenetrable
bosons on a lattice.

The paper is organised as follows: in Section~\ref{sec:obdm}
we use the Jordan-Wigner transformation to obtain the determinantal 
representation for the one-body density matrix of
impenetrable bosons. We discuss how  the results of Vaidya and Tracy
are obtained by considering the leading asymptotic
behaviour of the determinant. In Sections~\ref{sec:replica_g1} and 
\ref{sec:pert} we present the replica method and derive our main result, 
the full asymptotic expansion of the one-body density matrix. 
Section~\ref{sec:concl_prosp} contains conclusions and prospects.  
The mathematical details are given in Appendices.

\section{One-body density matrix as a Toeplitz determinant}
\label{sec:obdm}

Let  $a_m$, $a^\dagger_m$ be creation and annihilation operators  
of bosons on an infinite lattice with sites labeled by  the index $m$. 
In the limit of strong repulsive on-site interactions the Hilbert space is
projected onto a subspace of states with at most one particle in each
site. Using the Jordan-Wigner  transformation 
\begin{equation}
  \label{eq:JW_trans}
  a_m = \prod_{0<l<m} \!(-1)^{c^\dagger_l c_l}\; c_m;
  \qquad
  a^\dagger_m = c^\dagger_m \prod_{0<l<m} \!(-1)^{c^\dagger_l c_l}
\end{equation}
the bosonic operators  are related  
to fermionic creation and  annihilation operators, $c_m^\dagger$, $c_m$
obeying standard anti-commutation relations
\begin{equation}
  \label{eq:Fermi_comm}
 \{c_m,c^\dagger_l\}=\delta_{m,l};
 \qquad  \{c_m,c_l\}
 =\{c^\dagger_m,c^\dagger_l\}=0 .
\end{equation}

The one-body density matrix is defined by the following ground-state
expectation value:
\begin{eqnarray}
  g_1 (R) &=& 
  \frac{1}{2} \langle(a_{m+R}+a^\dagger_{m+R}) 
    (a_m+a^\dagger_m )\rangle \nonumber\\
  &=& \frac{1}{2} \langle(c_{m+R}+c^\dagger_{m+R}) \prod_{m<l<m+R} 
    \!(-1)^{c^\dagger_l c_l}\;(c_m-c^\dagger_m)\rangle,
  \label{eq:g1_def}
\end{eqnarray}
where we have used the relation
\begin{equation}
  \label{eq:the_fact}
  (-1)^{c^\dagger_l c_l} = (c_l-c^\dagger_l)(c_l+c^\dagger_l)
\end{equation}
and the fermionic commutation relations (\ref{eq:Fermi_comm}). 

For $R=0$ the one-body density matrix is related to the mean number of
particles per site (filling factor) $\nu$ as
\begin{equation}
  \label{eq:g1_0}
  g_1(0) =
  \langle c^\dagger_m c_m\rangle -\frac{1}{2} = \nu-\frac{1}{2},
\end{equation}
while for $R>1$ we use  the relation (\ref{eq:the_fact}) for every 
factor in the  product in Eq.~(\ref{eq:g1_def}) and  represent 
the one-body density matrix  as an expectation value of $2R$ operators:
\begin{equation}
  \label{eq:g1_AB}
  g_1 (R) = 
  \frac{1}{2} \langle A_{m+R} B_{m+R-1} A_{m+R-1} \ldots 
    B_{m+1} A_{m+1} B_m \rangle,
\end{equation}
where we have defined $A_l=c_l+c^\dagger_l$, $B_l=c_l-c^\dagger_l$.
We now use the Wick theorem and take into account that free fermionic 
expectation values  are given by
  \begin{eqnarray*}
    \langle A_{m+l} B_m\rangle &\equiv& G_l \\ 
    \langle A_{m+l} A_m\rangle &=&\langle B_{m+l} B_m\rangle = 0,
  \end{eqnarray*}
where the Green function of free fermions is
\begin{eqnarray}
  G_l &=& \langle 
  (c_{m+l}+c^\dagger_{m+l})(c_m-c^\dagger_m) \rangle =
  \int_{-\pi}^\pi \frac{dq}{2\pi}\, e^{iq l}\, (n_q+n_{-q}-1)  \nonumber \\ 
  &=& 
  \frac{2}{\pi}\frac{\sin \pi\nu l}{l} -\delta_{l,0}.
  \label{eq:Green}
\end{eqnarray}
Here $n_q = n_{-q}$ is the ground state momentum distribution of 
free fermions in a lattice with filling factor $\nu$, so that 
\begin{equation}
  \label{eq:nq}
n_q+n_{-q}-1  = \left\{
    \begin{array}{rl}
      1, & \qquad |q| < \pi \nu \\
      -1, & \qquad |q|>\pi\nu
    \end{array}\right. .
\end{equation}
Then the one-body density matrix (\ref{eq:g1_AB}) is represented as
a determinant of $R\times R$ matrix
\begin{equation}
  \label{eq:g1_det}
  g_1 (R) = \frac{1}{2}
  \left|
    \begin{array}{lllll}
      G_1     & G_2     & G_3     & \ldots  & G_R     \\
      G_0     & G_1     & G_2     & \ldots  & G_{R-1} \\
      G_{-1}  & G_0     & G_1     & \ldots  & G_{R-2} \\
      \vdots  & \vdots  & \vdots  & \ddots  & \vdots  \\
      G_{2-R} & G_{3-R} & G_{4-R} & \ldots  & G_1
    \end{array}
    \right| .
\end{equation}

The determinant (\ref{eq:g1_det}) has identical  elements 
along each of its diagonals and thus belongs to the class of  
Toeplitz determinants:
\begin{equation}
  \label{eq:Toeplitz}
  g_1 (R) = \det[g_{j-k}]_{j,k=1,\ldots , R};
  \qquad\qquad
  g_l = \frac{1}{2\pi} \int_{-\pi}^\pi  g(q) e^{iq l} dq,
\end{equation}
where $g(q)$ is called  generating function (symbol). It is easily shown from
(\ref{eq:Green}) that in our case the generating function is
\begin{equation}
  \label{eq:Toeplitz_gen}
  g (q) = e^{iq} (n_q+n_{-q}-1) = 
  \left(\frac{1-e^{i\pi\nu}e^{-iq}}{1-e^{-i\pi\nu}e^{iq}}\right)^\frac{1}{2}
  \left(\frac{1-e^{-i\pi\nu}e^{-iq}}{1-e^{i\pi\nu}e^{iq}}\right)^\frac{1}{2} .
\end{equation}
It  has two jump discontinuities at Fermi points $q=\pm\pi\nu$.  The
large size asymptotic behavior of Toeplitz determinants with such
singular generating functions was a subject of extensive studies in
mathematics (for overview and references see \cite{Basor1994}). It has
been shown for a certain class of singular generating functions that the
large size asymptotics of Toeplitz determinants is correctly described
by the Fisher-Hartwig conjecture \cite{Basor1994}.  In the  case of
generating function with two jumps as in Eq.~(\ref{eq:Toeplitz_gen}) the
analysis of the asymptotic behaviour of the Toeplitz determinant
(\ref{eq:g1_det}) is subtle and the Fisher-Hartwig asymptotic formula 
remains a conjecture. Being  applied, it  yields the following 
long-distance behaviour of the one-body density matrix:  
\begin{equation}
  \label{eq:g1_FH}
  g_1(R) = \frac{\rho_\infty}{\pi} 
  \left|\frac{\sin\pi\nu}{R}\right|^\frac{1}{2},\qquad R\to\infty,
\end{equation}
where $\rho_\infty = \pi e^{1/2}2^{-1/3} A^{-6}$, and
$A=\exp(1/12-\zeta'(-1))\simeq 1.2824271$ is Glaisher's constant related to
the  Rieman zeta function. Numerical computation of
the determinant (\ref{eq:g1_det}) for $R\sim 100$ and various values
of the filling factor shows that Eq. (\ref{eq:g1_FH}) provides an
excellent estimate for  the dominant smooth behaviour of the one-body
density matrix at large distances. Substracting this dominant
contribution from the exact numerical expression reveals
interesting oscillating corrections. These corrections are sensitive to
the specific value of the filling factor and reflect an interplay between
short-distance interparticle correlations and lattice
effects. The Fisher-Hartwig conjecture is unable to
capture this behaviour.  In the next Section we present the
calculations of the asymptotic long-distance behaviour of the one-body
density matrix, based on the replica method which has been recently
applied to study correlation properties of integrable models. We
shall see that this method is able to provide us with the dominant
behaviour (\ref{eq:g1_FH}) as well as with the oscillating terms.

\section{Replica calculations of the one-body density matrix}
\label{sec:replica_g1}

Below we describe the replica method for finding the large $R$ asymptotics  of
the determinant~(\ref{eq:g1_det}).
By standard manipulations the determinant in (\ref{eq:g1_det}) is
transformed to the following $R$-dimensional integral:
\begin{equation}
  \label{eq:g1_average}
  g_1 (R) = \frac{1}{2R!}\int_{-\pi}^\pi \frac{d^R q}{(2\pi)^R}
  |\Delta_R (e^{iq})|^2 \prod_{j=1}^R g(q_j) 
  \equiv \frac{1}{2}\left\langle\prod_{j=1}^R g(q_j)\right\rangle_R,
\end{equation}
where
\begin{equation}
  \label{eq:vdm}
  \Delta_R(e^{iq})=\Delta(e^{iq_1},e^{iq_2},\ldots,e^{iq_R})=
  \det[e^{i(k-1)q_l}]_{k,l=1,\ldots,R}=
  \prod_{1\le k<l\le R} (e^{iq_k}-e^{iq_l})
\end{equation}
is the Vandermonde determinant. The notation of average in
(\ref{eq:g1_average}) is  justified by the fact that the measure of
integration coincides with the distribution of eigenvalues of random
unitary matrices drawn from Dyson's Circular Unitary Ensemble
\cite{MehtaRandMatr}. Denoting $\exp(iq_l)=z_l$ and $\exp(i\pi\nu)=v$
and using the representation (\ref{eq:Toeplitz_gen}) we arrive at
an equation
\begin{eqnarray}
  g_1 (R) &=& \frac{1}{2}\left\langle\prod_{l=1}^R 
    \left(\frac{1-v\bar{z_l}}{1-\bar{v}z_l}\right)^\frac{1}{2}
    \left(\frac{1-\bar{v}\bar{z}_l}{1-vz_l}\right)^\frac{1}{2}
  \right\rangle_R \nonumber \\ 
  &=&
  \frac{1}{2}\left\langle\prod_{l=1}^R 
    \frac{|1-v\bar{z_l}|}{1-\bar{v}z_l}
    \frac{|1-\bar{v}\bar{z}_l|}{1-vz_l}
  \right\rangle_R
  \label{eq:g1_av1}
\end{eqnarray}

The replica evaluation of the average (\ref{eq:g1_average}) 
begins  with  modifying (replicating) the function being
averaged in (\ref{eq:g1_av1}) by taking the $2n$-th
power of  absolute values.  Then the original expression is recovered by
taking $n\to 1/2$. It is however assumed throughout the calculations
that $n$ is an \emph{integer} (and $2n$ is even) so that many
simplifications occur. Eventually,  the limit $n\to 1/2$ will be taken with
care and we shall describe it in detail.

For an integer $n$, a  straightforward
algebra shows that we are dealing with  an average of a rational
function: 
\begin{eqnarray}
  Z_n (R) &=&   \frac{1}{2}\left\langle\prod_{l=1}^R 
    \frac{|1-v\bar{z_l}|^{2n}}{1-\bar{v}z_l}
    \frac{|1-\bar{v}\bar{z}_l|^{2n}}{1-vz_l}
  \right\rangle_R \nonumber \\ 
  &=&
  \frac{v^{2nR}}{2}\left\langle\prod_{l=1}^R 
    \frac{(\bar{v}-\bar{z_l})^{2n}}{1-\bar{v}z_l}
    \frac{(\bar{v}-z_l)^{2n}}{1-vz_l}
  \right\rangle_R .
  \label{eq:g1_replica}
\end{eqnarray}
The crucial fact is the existance of a 
a duality transformation relating the $R$ - dimensional
integral (\ref{eq:g1_replica}) to a $2n-2=m$ - dimensional integral:
\begin{equation}
  \label{eq:duality}
  Z_n (R) =  \frac{(-1)^{m+1}}{2S_m} 
  \int_0^1 d^m x \Delta^2_m (x) 
  \prod_{c=1}^m x_c (1-x_c) \left[(1-x_c)v+\bar{v}x_c\right]^R,
\end{equation}
where 
\begin{equation}
  \label{eq:Sm}
  S_m=m!\prod_{c=1}^m\frac{\Gamma^2(1+c)\Gamma(R+m+1-c)}
  {\Gamma(R+m+2+c)} 
\end{equation}
The proof of the duality transformation is presented in Appendix
\ref{app:duality}. 

Note that in the dual representation (\ref{eq:duality}) large
distance $R$ appears only as a parameter, which makes the dual 
representation  an excellent 
starting point for the asymptotic expansion of $g_1(R)$. The main
contribution to the integral (\ref{eq:duality}) comes from the
end-points of integration.  Let $p$ be a number of
variables in the vicinity of $x_-=0$ and $p'=m-p$ be a number of
variables close to $x_+=1$. We shift the integration variables and
approximate the integrated high-degree polynomial  as 
\begin{eqnarray}
  x_c&=&x_-+\frac{\xi_a}{R(1-v^{-2})},\qquad
  \left[(1-x_c)v+\bar{v}x_c\right]^R \simeq v^R e^{-\xi_c},\qquad
  c=1,\ldots,p \nonumber \\
  x_{p+d}&=&x_+-\frac{\xi'_d}{R(1-v^2)},\qquad
  \left[(1-x_{p+d})v+\bar{v}x_{p+d}\right]^R \simeq v^{-R} e^{-\xi'_d},\qquad
  d=1,\ldots,p'
  \label{eq:stationary}
\end{eqnarray}
The integration measure in (\ref{eq:duality}) 
factorizes as
\begin{eqnarray}
  d^m x\; \Delta^2_m (x) 
  \prod_{c=1}^{m} x_c (1-x_c) &\simeq&
  \left[\frac{1}{R(1-v^{-2})}\right]^{p(p+1)} 
  d^p \xi_c\; \Delta^2_p (\xi_c)\prod_{c=1}^p \xi_a \nonumber \\
  &\times&\left[\frac{1}{R(1-v^2)}\right]^{p'(p'+1)}
    d^{p'} \xi'_d \;\Delta^2_{p'} (\xi'_d)\prod_{d=1}^{p'} \xi'_d .
  \label{eq:factor_measure}
\end{eqnarray}
Then, summing over all possibilities to distribute $m$ variables among
saddle points and performing the remaining integration over $\xi_c$,
$\xi'_d$ we obtain 
\begin{equation}
  \label{eq:Zn_series1}
  Z_n (R) = \frac{(-1)^{m+1}}{S_m}\sum_{\substack{p,p' \\ p+p'=m}} 
  \frac{m!}{p!p'!}
  I_p I_{p'}
  \left[\frac{1}{R(1-v^{-2})}\right]^{p(p+1)}
  \left[\frac{1}{R(1-v^2)}\right]^{p'(p'+1)} v^{R(p-p')}, 
\end{equation}
where the integrals are performed using the Selberg integration formula
\begin{equation}
  \label{eq:selberg}
  I_p = \int_0^\infty d^p\xi\;\Delta_p^2 (\xi) \prod_{c=1}^p \xi_c
  e^{-\xi_c} =\prod_{c=1}^p \Gamma^2(1+c).
\end{equation}
Combining them with the constant (\ref{eq:Sm}) we observe that each
term in the sum (\ref{eq:Zn_series1}) is proportional to the total
combinatorial factor:
\begin{equation}
  \label{eq:comb_total}
  \frac{\Gamma^2(m+3)}{\Gamma(p+1)\Gamma(p'+1)}\left[F^{p+1}_{m+2}\right]^2
  \prod_{c=1}^m\frac{\Gamma(R+m+2+c)}{\Gamma(R+m+1-c)},
\end{equation}
where
\begin{equation}
  \label{eq:Fnl}
  F^{p}_{m}=\frac{\prod_{d=1}^p \Gamma(d) \prod_{b=1}^{m-p} \Gamma
  (b)}{\prod_{c=1}^m \Gamma(c)}
\end{equation}
is a familiar combination from previous studies of replica theories
\cite{KamenevMezard1999,GangardtKamenev2001}.

Now we are at a position to take the limit $n\to 1/2$. It is important to
notice that for even $m=2n-2$ there is always a central term $p=p'=m/2$ in the
sum (\ref{eq:Zn_series1}) which provides a smooth non-oscillatory
contribution.  It is characterized by a maximal degree of replica symmetry
breaking, since it originates from the saddle point for which the number of
components $x_d\sim x_-$ is equal to the number of components $x_b\sim x_+$.
This term will be retained in the limit $n\to 1/2$ ($m\to -1$) and this
provides an operational definition of the correct analytic continuation.

Making a change of summation index from $p$ to $m/2+k=n-1+k$ we factorize
the combinatorial coefficient~(\ref{eq:Fnl}) as 
$F^{n+k}_{2n} = A_n D^{(n)}_k$, where 
\begin{eqnarray}
  \label{eq:Anl}
  A_n&=&\prod_{c=1}^n\frac{\Gamma(c)}{\Gamma(2n+1-c)},\\
  \label{eq:Dnl}
  D^{(n)}_k &=&\prod_{c=1}^k \frac{\Gamma(n+c)}{\Gamma(n+1-c)}=
  \frac{\Gamma(n+c)}{\Gamma(n+1-c)} D^{(n)}_{k-1},\qquad k>0;\qquad
  D^{(n)}_0=1;\qquad D^{(n)}_k=D^{(n)}_{-k} .
\end{eqnarray}
This leads to an expression 
\begin{equation}
  \label{eq:Zn_series2}
  Z_n (R) = \frac{C_n (R) A^2_n}{2} 
  \left|\frac{1}{R (\bar{v}-v)}\right|^{2n(n-1)}
  \sum_{k=-\infty}^\infty 
  \frac{\left[D^{(n)}_k\right]^2}{\Gamma(n+k)\Gamma(n-k)}
  \left|\frac{1}{ R (\bar{v}-v) }\right|^{2k^2}
  \left(\frac{1-v^2}{1-\bar{v}^2}\right)^{(2n-1)k}.
\end{equation}
Here we have defined
\begin{equation}
  \label{eq:Cm}
  C_n (R)= (-1)^{2n-1}\Gamma^2(2n+1)\prod_{c=1}^{2n-2}
  \frac{\Gamma(R+2n+c)}{\Gamma(R+2n-1-c)}
\end{equation}
and the summation over $k$ in Eq.~(\ref{eq:Zn_series2}) was extended to all
integers.

Now we take the limit $n\to 1/2$ term by term in the series
(\ref{eq:Zn_series2}).  The only complication appears with the overall
numerical factors $C_n$ and $A_n$. The analytical continuation of
these factors is explained in Appendix~\ref{app:analAn} where we show that 
$A^2_\frac{1}{2}=\sqrt{2} \rho_\infty$ and $C_\frac{1}{2}=1/R$. We
also have $\Gamma(1/2+k)\Gamma(1/2-k)=\pi(-1)^k$. We therefore find
that
\begin{equation}
  \label{eq:Z_1/2_series1}
  g_1(R)=Z_\frac{1}{2}(R) = \frac{\rho_\infty}{\pi} 
  \left|\frac{\sin\pi\nu}{R}\right|^\frac{1}{2}\left(1+
    2\sum_{k=1}^\infty (-1)^k  \left[D^{(\frac{1}{2})}_k\right]^2 
    \frac{\cos 2 \pi k \nu R}{\left(2 R\sin \pi\nu\right)^{2k^2}}\right).
\end{equation}

The first several terms in this  asymptotic expansion of the one-body 
density matrix read 
\begin{equation}
  \label{eq:g_1_series2}
  g_1(R) = \frac{\rho_\infty}{\pi} 
  \left|\frac{\sin\pi\nu}{R}\right|^\frac{1}{2}\left(1-  
    \frac{1}{8 \sin^2 \pi\nu} 
    \frac{\cos 2 \pi  \nu R}{ R^2}+
    \frac{9}{32768 \sin^8 \pi\nu} 
    \frac{\cos 4 \pi  \nu R}{ R^8}+\ldots\right),
\end{equation}
where the leading term coincides with the result~(\ref{eq:g1_FH}).  
As is seen from Eq.~(\ref{eq:g_1_series2}), for small $k$ the coefficients 
in the series (\ref{eq:Z_1/2_series1}) rapidly decrease with increasing $k$. 
However, they will diverge for large $k$, which shows that we are dealing here with
asymptotic, rather than convergent series.

\section{Perturbation Theory}
\label{sec:pert}

In Eq.~(\ref{eq:g_1_series2}),  only the leading contribution of
each saddle point was taken into account. To obtain the subleading
terms one has to construct the perturbation  theory around each saddle
point. Using the shift of  variables (\ref{eq:stationary}) the integrand in
(\ref{eq:duality}) is expanded in series in $1/R$.
As a result, using the definition of the average  
\begin{equation}
  \label{eq:pert_average}
  \Big\langle f(\xi)g(\xi')\Big\rangle_{p,p'}=
  \Big\langle f(\xi)\Big\rangle_p\Big\langle g(\xi')\Big\rangle_{p'}
\qquad\qquad
\Big\langle f(\xi)\Big\rangle_p=
  \frac{1}{I_p}
  \int_0^\infty d^p\xi\,f(\xi) \Delta_p^2(\xi)\prod_{c=1}^p \xi_c e^{-\xi_c}  
\end{equation}
each term in the  series (\ref{eq:Zn_series1}) gets multiplied by 
\begin{eqnarray}
  \label{eq:pert_f_1}
  F_{p,p'} (R) &=&\Big\langle
  \prod_{c=1}^p\prod_{d=1}^{p'}
  \left(1-\frac{\xi_c}{R(1-\bar{v}^2)}-\frac{\xi'_d}{R(1-v{^2})}\right)
  \prod_{c=1}^p \left(1-\frac{\xi_c}{R(1-\bar{v}^2)}\right)
  \prod_{d=1}^{p'}\left(1-\frac{\xi'_d}{R(1-v{^2})}\right)\nonumber\\
  &\times& \prod_{c=1}^p e^{-\xi_c^2/2R-\xi_c^3/3R^2-\ldots} 
  \prod_{d=1}^{p'} e^{-{\xi'}_d^2/2R-{\xi'}_d^3/3R^2-\ldots}
  \Big\rangle_{p,p'}
\end{eqnarray}

Up to terms of the order of $1/R^2$, it is sufficient to put $p=m/2=-1/2$,
since the leading contribution of other saddle points is of the order of
$1/R^2$ or higher. The function $F_{m/2,m/2} (R)$ is then evaluated using the
following averages (see chapter 17 in
\cite{MehtaRandMatr}) :
\begin{eqnarray}
&&\Big\langle \xi_1\Big\rangle_p = p+1,
\nonumber\\
\nonumber\\
&&\Big\langle \xi_1^2\Big\rangle_p = (2p+1)(p+1),
\qquad\qquad\qquad\qquad\;\;\;\;\;
\Big\langle \xi_1\xi_2\Big\rangle_p =p(p+1)
\nonumber\\
\nonumber\\
&&\Big\langle \xi_1^3\Big\rangle_p = (p+1)(5p^2+5p+2),
\qquad\qquad\qquad\;\;\;
\Big\langle \xi_1^2\xi_2\Big\rangle_p = 2p^2(p+1)
\nonumber\\
\nonumber\\
&&\Big\langle \xi_1^4\Big\rangle_p=(2p+1)(p+1)(7p^2+7p+2)
\qquad\qquad
\Big\langle \xi_1^2\xi_2^2\Big\rangle_p=2p^2(2p-1)(p+1) .
\label{eq:pert_mehta_av}
\end{eqnarray}
The result then  is
\begin{equation}
  \label{eq:pert_F_res}
  F_{-1/2,-1/2}(R)=1-\frac{\ctg^2\pi\nu}{32 R^2}
\end{equation}
which leads to the following asymptotic expression for the one-body
density matrix:
\begin{equation}
  \label{eq:g_1_final}
  g_1(R) = \frac{\rho_\infty}{\pi} 
  \left|\frac{\sin\pi\nu}{R}\right|^\frac{1}{2}\left(1-
    \frac{\ctg^2\pi\nu}{32 R^2}-
    \frac{1}{8 \sin^2 \pi\nu} 
    \frac{\cos 2 \pi  \nu R}{ R^2} + \ldots\right)
\end{equation}

Our result 
reproduces correctly the continuous limit. In this case it is useful to
introduce a length scale, the lattice constant $a$, so that $x=Ra$ is
the distance and $\pi\nu/a=k_F$ is the Fermi wavenumber proportional
to the density. Going to the continuous limit corresponds to having
the filling factor $\nu$ tending to zero while the product $k_F x = \pi\nu R$ is kept fixed. Dividing $g_1$ by the lattice constant to have a correct 
normalization, Eq.~(\ref{eq:g_1_series2}) becomes
\begin{equation}
  \label{eq:g1_continuum}
  \tilde{g}_1 (x) = (1/a) g_1 ( x/a )= \frac{\rho_\infty}{|k_F x|^\frac{1}{2}} 
  \left(1- \frac{1}{32 (k_F x)^2}- \frac{1}{8} \frac{\cos 2k_F x}{(k_F x)^2 }+\ldots\right)
\end{equation}
in agreement with the Vaidya and Tracy result \cite{VaidyaTracy1979}. The
sign  of the oscillating cosine term in Eq.~(\ref{eq:g1_continuum}) 
has been corrected as discussed in \cite{Gangardt2004}.

\section{Conclusions and prospects}
\label{sec:concl_prosp}

The result (\ref{eq:g_1_final}) provides an exact analyical expression for the
long distance behaviour of the one-body density matrix for impenetrable bosons
on an infinite lattice. It is valid for all values of the filling factor
$\nu$, in particular it reproduces correctly the continuous limit $\nu\to
0$. The structure of Eq.~(\ref{eq:g_1_final}) is in accordance with the
hydrodynamic expansion conjectured by Haldane
\cite{Haldane1981}. Equations~(\ref{eq:Anl}), and (\ref{eq:Dnl}) (at $n=1/2$)
provide the exact values for the non-universal coeficients of the leading
oscillating terms.

The oscillatory terms in (\ref{eq:g_1_final}) reflect the physics on short
distance scales, of the order of the mean interparticle separation. They are
analogous to the Friedel oscillations in the physics of fermions and provide
yet another manifestation of fermionization. The interplay between these
short-distance correlations and lattice effects is interesting, for instance,
in the case of half filling ($\nu =1/2$), where the particles tend to occupy
every second site. This results in the oscillations of the one-body density
matrix with the period of the lattice.

The momentum distribution cannot be reconstructed from the asymptotic
expression (\ref{eq:g_1_final}) for the one-body density matrix, as it
requires the knowledge of the short-distance behaviour. However, we can
predict some of the important features.  For sufficiently small momenta their
distribution diverges as an inverse square root of the momentum. The
finite size would result in the finite peak value of the momentum distribution
proportional to the square root of the total number of particles.  For momenta
of the order of $2k_F=2\pi \nu/a$ the derivative of the momentum distribution
has a jump (the distribution itself has a cusp) resulting from the underlying Fermi surface of the fermionized
bosons. Weaker singularities exist for higher multiples of $2k_F$. It is again
interesting to see that for the case of half filling the jump in the
derivative occurs just at the border of the Brillouin zone and therefore
cannot be observed: the momentum distribution has no singularities. 

These features can be probed in current experiments with cold atoms. 
The most promising method is expected to be a combination of Bragg spectroscopy
with the time of flight technique. In this case a Bragg pulse is applied after
switching off the trap, when the density of the cloud is already
sufficiently small so that  the number of scattered atoms (proportinal to the
momentum disitribution) can be  detected with precision \cite{Richard2003}. 
This technique can allow the observation of the singularities such as 
cusps in the momentum  distribution originating from the oscillating 
subleading terms in the coordinate correlation functions.

There are several directions in which the present work can be extended. An
immediate question concerns the finite value of the interactions or, in other
words the case of the soft-core bosons.  This case is similar to the case of
the Heisenberg spin chain, for which there are recent results based on the
Bethe Ansatz solution \cite{SpinChainCorr}.  It is not clear at the moment how
to obtain the explicit behaviour of the correlation functions away from the
free fermionic point (equivalent to the hard-core bosons) where the Toeplitz
determinant expression exists for the one-body density matrix. In the case of
impenetrable bosons, our work can be extended to the domain of time-dependent
correlation functions which contains information about elementary excitations
and are related to the responce of the system to external perturbations.

\section{Acknowledgments}
\label{sec:acknowledgments}

D.M.G would like to acknowledge the hospitality of the Physics Department of
the University of Minnessota and fruitful discussions with A. Kamenev. This
work was supported by the Minist\`ere de la Recherche (grant ACI Nanoscience
201), by the ANR (grants NT05-2\_42103 and 05-Nano-008-02), and by IFRAF
Institute. The research has been performed in part at the Van der Waals -
Zeeman Institute of the University of Amsterdam 
(supported by the Nederlandse Stichtung voor Fundamenteel
Onderzoek der Materie, FOM)  and  at KITP at Santa Barbara 
(supported by the National Science Foundation under 
Grant No. PHY99-07949).  LPTMS is a mixed research unit No. 8626 of 
CNRS and Universit\'e Paris Sud

\appendix

\section{Duality transformation}
\label{app:duality}

The average (\ref{eq:g1_replica}) is symmetric under the interchange of
$v$ and $\bar{v}$, so that using the fact that $z^{-1}_l=\bar{z}_l$ it is
possible to bring Eq.~(\ref{eq:g1_replica}) to the form
\begin{eqnarray}
  Z_{n} (R) &=&  \frac{(-1)^R}{2}
\left\langle\prod_{l=1}^R \bar{z}_l(1-\bar{v}\bar{z_l})^{2n-1}
  (v-z_l)^{2n-1}
  \right\rangle   \nonumber \\
  &=&\lim_{\substack{v_1\to\infty,\;
      v_2,\ldots,v_{2n}\to v\\
      \bar{u}_1,\ldots,\bar{u}_{2n-1}\to v,\;\bar{u}_{2n}\to 0}}
  \frac{\prod_{c=1}^{2n}v_c^{-R}}{2}
\left\langle\prod_{l=1}^R \prod_{c=1}^{2n}(\bar{u}_c-\bar{z_l})
    (v_c-z_l)
    \right\rangle,
  \label{eq:Zn_replica}
\end{eqnarray}
where we have ``split'' $2(2n-1)$ points in the product using
variables $\bar{u}_c$ for $c=1,\ldots, 2n-1$  and  $v_c$ for
$c=2,\ldots,2n$.  We have also defined two
additional variables $v_1$ and $\bar{u}_{2n}$ in order to represent
the factor $z_l$ in the product on the same footing. 

Recalling the definition (\ref{eq:g1_average}) for the average and
using an obvious relation
\begin{equation}
  \label{eq:vdm_prod}
  \prod_{l=1}^R\prod_{c=1}^{2n} (v_c-z_l) =
  \frac{\Delta_{R+2n} (v_1,\ldots,v_{2n}, z_1,\ldots,z_R)}
  {\Delta_{2n}(v_1,\ldots,v_{2n})
    \Delta_{R} (z_1,\ldots,z_R)}
\end{equation}
we  rewrite Eq.~(\ref{eq:g1_replica}) as a many-body correlation function
\begin{eqnarray}
  Z_{n} (R)&=&  \frac{\prod_{c=1}^{2n}v_c^{-R}}{2
    \Delta_{2n}(\bar{u})\Delta_{2n}(v)}\;
 \frac{1}{R!}\int_{-\pi}^\pi \frac{d^R q}{(2\pi)^R}
  \Delta_{R+2n} (\bar{u}_1,\ldots,\bar{u}_{2n}, \bar{z}_1,\ldots,\bar{z}_R)
  \Delta_{R+2n} (v_1,\ldots,v_{2n},
  z_1,\ldots,z_R)\nonumber\\
  &=&
    \frac{\prod_{c=1}^{2n}v_c^{-R}}{2\Delta_{2n}(\bar{u})
      \Delta_{2n}(v)}
  \langle 0|\psi^\dagger
 (u_1)\ldots\psi^\dagger(u_{2n})
 \psi(v_{2n})\ldots \psi(v_1)|0\rangle
  \label{eq:g1_manybody}
\end{eqnarray}
of fermionic creation and annihilation operators $\psi^\dagger (\bar{s}_c)$,
$\psi (\bar{t}_c)$ in the ground state of $R+2n$ fermions.
Up to normalization, the ground state wavefunction is given by the 
Vandermonde determinants: $\langle z|0\rangle\propto\Delta_{R+2n}(z)$. 
The expectation value of the fermionic operators is given by 
the Wick theorem in the form of $2n\times 2n$ determinant:
\begin{eqnarray}
  \langle 0|\psi^\dagger
 (u_1)\ldots\psi^\dagger(u_{2n})
 \psi(v_{2n})\ldots \psi(v_1)|0\rangle=
\left|
    \begin{array}{ccccc}
      G(\bar{u}_1 v_1)   & G(\bar{u}_1 v_2)   & G(\bar{u}_1 v_3) & \ldots  & G(\bar{u}_1 v_{2n})\\
      G(\bar{u}_2 v_1)   & G(\bar{u}_2 v_2)   & G(\bar{u}_2 v_3) & \ldots  & G(\bar{u}_2 v_{2n})\\
      G(\bar{u}_3 v_1)   & G(\bar{u}_3 v_2)   & G(\bar{u}_3 v_3) & \ldots  & G(\bar{u}_3 v_{2n})\\
      \vdots  & \vdots  & \vdots  & \ddots  & \vdots  \\
      G(\bar{u}_{2n} v_1)   & G(\bar{u}_{2n} v_2)   & G(\bar{u}_{2n} v_3) &
      \ldots  & G(\bar{u}_{2n} v_{2n})     \\
    \end{array}
    \right| 
  \label{eq:Wick}
\end{eqnarray}
with the following 2-fermionic Green function:
\begin{equation}
  \label{eq:fermiG}
  G(\bar{u} v)=\langle 0|\psi^\dagger (u)\psi(v)|0\rangle=\sum_{p=0}^{R+2n-1} \bar{u}^p v^p = \frac{1-(\bar{u}v)^{R+2n}}{1-\bar{u}v}.
\end{equation}

We now take the limits $\bar{u}_{2n}\to 0$,  
$v_1\to\infty$ using $G(\bar{u}_{2n}v)=G(0)=1$ and
$G(\bar{u} v_1)\simeq (\bar{u} v_1)^{R+2n-1}$. The factor
$v_1^{R+2n-1}$ common to the elements in the first column cancels 
with the corresponding term in the prefactor of the second line of
Eq.~(\ref{eq:Zn_replica}).  
Taking out a factor $\bar{u}_c^{R+2n-1}$ from each row we arrive
at the following representation:
\begin{equation}
  \label{eq:Zn_1}
  Z_{n} (R)  = -\frac{1}{2}\frac{\prod_{c=1}^{2n-1} \bar{u}_c^{R} v_{c+1}^{-R}}
  {\Delta_{2n-1} (u)\Delta_{2n-1} (v)}  
  \left|
    \begin{array}{ccccc}
      1   & \widetilde{G}(u_1, v_2)   & \widetilde{G}(u_1, v_3) & \ldots  & \widetilde{G}(u_1, v_{2n})     \\
      1   & \widetilde{G}(u_2, v_2)   & \widetilde{G}(u_2, v_3) & \ldots  & \widetilde{G}(u_2, v_{2n})     \\
      \vdots  & \vdots  & \vdots   & \ddots  & \vdots \\
      1   & \widetilde{G}(u_{2n-1}, v_2)   & \widetilde{G}(u_{2n-1}, v_3) & \ldots  & \widetilde{G}(u_{2n-1}, v_{2n}) \\
      0                     & 1                 & 1               & \ldots  & 1   \\
    \end{array}
    \right| ,
\end{equation}
where $\widetilde{G}(u,v)=(u^{R+2n}-v^{R+2n})/(u-v)$
and we have used the relation $\Delta_m (\bar{u}_1,\ldots,\bar{u}_m) = -
\prod_{a=1}^m \bar{u}^{m-1}_a\Delta_m(u_1,\ldots,u_m)$.

We take the singular limit  $u_c\to u$ and $v_c\to v$  by a standard procedure,
described, for example in \cite{MehtaMatr}. Expanding in non-zero elements of 
the last row and first column 
leads to  expressing Eq.~(\ref{eq:Zn_1}) as a 
 determinant of  $(2n-2)\times (2n-2)$ matrix of  partial derivatives:
\begin{equation}
  \label{eq:Zn_2}
  Z_{n} (R)  = \frac{(-1)^{2n-1}}
  {2\prod_{c=1}^{2n-1}\Gamma^2 (c)}  
  \left|
    \begin{array}{cccc}
      \partial_u\partial_v \widetilde{G}   
      & \partial_u\partial^2_v \widetilde{G}  &
      \ldots  & \partial_u\partial^{2n-2}_v \widetilde{G}     \\
      \partial^2_u\partial_v \widetilde{G}   
      & \partial^2_u\partial^2_u \widetilde{G}  &
      \ldots  & \partial^2_u\partial^{2n-2}_u \widetilde{G}     \\
      \vdots   & \vdots   & \ddots  & \vdots \\
      \partial^{2n-2}_u\partial_v \widetilde{G}   
      & \partial^{2n-2}_u\partial^2_v \widetilde{G}  &
      \ldots  & \partial^{2n-2}_u\partial^{2n-2}_v \widetilde{G}  
    \end{array}\right|
\end{equation}

Now we use the following integral representation in which we recognize
the expression for the hypergeometric function:
\begin{eqnarray}
  \partial_u\partial_v \widetilde{G}(u,v) &=&
  \frac{\Gamma(R+2n+1)}{\Gamma(R+2n-2)}\int_0^1 \!dx\,
  x(1-x)[(1-x)u+vx]^{R+2n-3}\nonumber\\
  &=&\frac{\Gamma(R+2n+1)}{\Gamma(R+2n-2)}\frac{\Gamma^2 (2)}{\Gamma(4)}
  \,u^{R+2n-3}F (-R-2n+3,2;4;s),
  \label{eq:int_rep}
\end{eqnarray}
where $s=1-\bar{u}v$. Splitting again the  variables $s\to s_c=1-\bar{u}_c
v$, for $c=1,\ldots,2n-2$ the determinant
(\ref{eq:Zn_2}) is represented once again as a ratio
\begin{eqnarray}
  Z_{n} (R)  &=& \frac{(-1)^{2n-1}}
  {2\Gamma^2(2n-1)\prod_{c=1}^{2n-2}\Gamma (c)}
  \left[\frac{\Gamma(R+2n+1)}{\Gamma(R+2n-2)}\frac{\Gamma^2
      (2)}{\Gamma(4)}\right]^{2n-2}\frac{\prod_{c=1}^{2n-2}u_c^R}{\Delta_{2n-2}(s)}
\nonumber\\ &\times&
  \left|
    \begin{array}{cccc}
       F_0 (s_1)   &  F_1(s_1)  &  \ldots  &    F_{2n-3}(s_1)     \\
       F_0 (s_2)   &  F_1(s_2)  &  \ldots  &    F_{2n-3}(s_2)     \\
      \vdots   & \vdots   & \ddots  & \vdots    \\
       F_0(s_{2n-2})   & F_1(s_{2n-2})  &  \ldots  &  F_{2n-3}(s_{2n-2})     
    \end{array}\right|,
  \label{eq:Zn_3}
\end{eqnarray}
where $F_k(s)=[(s-1)\partial_s]^k F(-R-2n+3,2;4;s)$. In this function $k$ derivatives
$\partial_s$ can be commuted through the factors $s-1$ to the right,
producing additional terms which has at most $k-1$ derivatives. These
terms do not contribute to the determinant, since they already appear in the
$k-1$-th column. Next, we use the known properties of hypergeometric
functions to show that
\begin{eqnarray*}
  \label{eq:F_trans}
  F_k(s)&=&(s-1)^k\partial_s^k F(-R-2n+3,2,4,s)= \frac{(-R-2n+3)_k
    (2)_k}{(4)_k} (s-1)^k  F(-R-2n+3+k,2+k;4+k;s) \\
  &=&
  \frac{\Gamma(R+2n-2)}{\Gamma(R+2n-2-k)}
  \frac{\Gamma(2+k)}{\Gamma(2)}\frac{\Gamma(4)}{\Gamma(4+k)}(1-s)^{R+2n-1}
  F(R+2n+1,2;4+k;z). 
\end{eqnarray*}

Substituting this expression into (\ref{eq:Zn_3}) and taking the limit
$s_c\to s=1-\bar{u}v$ we get
\begin{eqnarray}
  Z_{n} (R)  &=& \frac{(-1)^{2n-1}}{2}
  \prod_{c=1}^{2n-2}\frac{\Gamma(2)\Gamma(R+2n+1)}
  {\Gamma(1+c)\Gamma(3+c)\Gamma(R+2n-1-c)}  (1-s)^{(n-1)(R+4n-2)}
\nonumber\\ &\times&
  \left|
    \begin{array}{cccc}
       \widetilde{F}(0;s)   & \widetilde{F} (1;s)   &  \ldots  &
       \widetilde{F} (2n-3;s)  \\
       \partial_s \widetilde{F} (0;s)   & \partial_s \widetilde{F} (1;s)    
       &  \ldots  & \partial_s \widetilde{F} (2n-3;s)         \\
      \vdots   & \vdots   & \ddots  & \vdots    \\
       \partial^{2n-3}_s \widetilde{F} (0;s)   & \partial^{2n-3}_s
       \widetilde{F} (1 ;s) 
       &  \ldots  &  \partial^{2n-3}_s \widetilde{F}(2n-3; s)     
    \end{array}\right|,
  \label{eq:Zn_4}
\end{eqnarray}
where $\widetilde{F}(p;s) = F(R+2n+1,2;4+p;s)$.  For the
derivatives  of the function $\widetilde{F}_p (s)$ we have  
\begin{eqnarray*}
  \label{eq:Ftilde}
  \partial^k_s \widetilde{F}(p;s) &=&\frac{(N+2n+1)_k (2)_k }{(4+p)_k}
  F(N+2n+1+k,2+k;4+p+k;s)  \\
  &=&\frac{(R+2n+1)_k (2)_k }{(4+p)_k}
    (1-s)^{-R-2n+1+p-k} F(-R-2n+3+p,2+p;4+p+k;s)\\
  &=&\frac{\Gamma(R+2n+1+k)}{\Gamma(R+2n+1)}
  \frac{\Gamma(4+p)}{\Gamma(2)\Gamma(2+p)}(1-s)^{-R-2n+1+p-k} 
\int_0^1 dx\,x^{p+1} (1-x)^{k+1} (1-s x)^{R+2n-3-p}.
\end{eqnarray*}
In this expression we used the properties of hypergeometric
functions \cite{GradshteynRyzhik} along with its integral representation. 
Substituting this result into (\ref{eq:Zn_4}) yields
\begin{eqnarray}
  Z_{n} (R)  &=& \frac{(-1)^{2n-1}}{2}
  \prod_{c=1}^{2n-2}\frac{\Gamma(R+2n+c)}
  {\Gamma^2(1+c)\Gamma(R+2n-1-c)} (1-s)^{R(n-1)}
  \nonumber\\
  &\times&\int_0^1 dx_1\ldots dx_{2n-2} \;
  (1-x_1)^0 (1-x_2)^1 \ldots (1-x_{2n-2})^{2n-3} \prod_{c=1}^{2n-2}
  x_c(1-x_c)
  \nonumber\\ &\times&
  \left|
    \begin{array}{cccc}
      (1-s x_1)^{R+2n-3}     & x_1 (1-s x_1)^{R+2n-4}      &
      \ldots  & x_1^{2n-3}(1-s x_1)^R     \\
      (1-s x_2)^{R+2n-3}      
      & x_2 (1-s x_2)^{R+2n-4}      &  \ldots  & 
      x_2^{2n-3} (1-s x_2)^R       \\
      \vdots   & \vdots   & \ddots  & \vdots    \\
      (1-s x_{2n-2})^{R+2n-3}     
      &  x_{2n-2}(1-s x_{2n-2})^{R+2n-4}
      &  \ldots  & x_{2n-2}^{2n-3}(1-s x_{2n-2})^R       
    \end{array}\right|.
  \label{eq:Zn_5}
\end{eqnarray}
The first factor in the integrand  is
anti-symmetrized to yield Vandermonde determinant $\Delta_{2n-2} (1-x)
= -\Delta_{2n-2} (x)$.  The
integration can be performed column-wise. This enables one to extract another factor 
$-\Delta_{2n-2} (x)$. Finally, setting $m=2n-2$ and $1-s=v^2$  and
interchanging $v$ and $\bar{v}$ we
obtain Eq.~(\ref{eq:duality}).

\section{Analytical continuation of products $A_n$ and $C_n (R)$}
\label{app:analAn}

The main idea behind evaluating the products $\prod_{c=1}^n f_c $
such as  (\ref{eq:Anl}) and (\ref{eq:Cm}) 
for fractional or negative values of the product limit $n$
is to turn the products into exponentials of  sums of logarithms of
each factor and then use the  integral representation 
\cite{GradshteynRyzhik}  for the logarithm of 
Euler's gamma function. 
It is possible to perform the summation  under the integral and
obtain an integral representation  where $n$ enters only as a parameter. 
The resulting integral representation can be
analytically continued to a desired value of $n$. 
In the case of $A_n$ the procedure is described in detail in Appendix~B of
Ref.~\cite{Gangardt2004}. We cite here the final result: 
\begin{equation}
 \rho_\infty=A^2_{1/2}/\sqrt{2} = \pi e^{1/2} 2^{-1/3} A^{-6}.
\end{equation}

For the function $C_n (R)$ defined by Eq.~(\ref{eq:Cm}) we have
\begin{eqnarray}
  \ln \frac{C_n (R)}{(-1)^{2n-1} \Gamma^2(2n+1)}&=&
  \sum_{c=1}^{2n-2}\left(\ln \Gamma(R+2n+c) -
    \ln \Gamma(R+2n-1-c)\right) \nonumber\\
  &=&\int_0^\infty\frac{dt}{t}\left[
      e^{-Rt}\sum_{c=1}^{2n-2}\frac{e^{-(2n+c)t}-e^{-(2n-1-c)t}}{1-e^{-t}}
    +\sum_{c=1}^{2n-2} (2c+1)e^{-t}\right] \nonumber\\
  &=& \int_0^\infty\frac{dt}{t}\left[
      e^{-Rt}\frac{e^{-(2n+1)t}-e^{-(4n-1)t}+e^{-(2n-1)t} -e^{-t}}
      {\left(1-e^{-t}\right)^2}
    +2n(2n-2)e^{-t}\right]
\end{eqnarray}
For $n=1/2$ the first term in square brackets is exactly equal to $e^{-Rt}$,
and we obtain the desired analytical continuation 
\begin{equation}
  \label{eq:C1/2}
  \ln C_\frac{1}{2}(R) = \int_0^\infty\frac{dt}{t}\left(
      e^{-Rt}-e^{-t}\right) = \ln \frac{1}{R}.
\end{equation}
Thus $C_\frac{1}{2} (R)=1/R$.


\vspace{-0.5cm}


\begin{thebibliography}{99}


\bibitem{KorepinBook} V.E. Korepin, N.M. Bogoliubov and A.G. Izergin,
  \textit{Quantum Inverse Scattering Method and Correlation
  Functions}, (Cambridge University Press 1993).


\bibitem{Laburthe2004} B.~Laburthe Tolra, K.M.~O'Hara, J.H.~Huckans, 
W.D.~Phillips, S.L.~Rolston, and J.V.~Porto, Phys. Rev. Lett.
\textbf{92}, 190401 (2004). 

\bibitem{Kinoshita2004} T.~Kinoshita, T.Wenger, and D.S.~Weiss,
  Science \textbf{305}, 1125 (2004).



\bibitem{Paredes2004} B.~Paredes, A.~Widera, V.~Murg, O.~Mandel,
  S.~F\"olling, I.~Cirac, G.V.~Shlyapnikov, T.W.~H\"ansch, I.~Bloch,
  Nature \textbf{429}, 277, (2004). 


\bibitem{Koehl2004} M.~K\"ohl, T.~St\"oferle, H.~Moritz, C.~Schori, and
  T.~Esslinger, App. Phys. B \textbf{79}, 1009 (2004).

\bibitem{Fertig2005} C.D.~Fertig,
  K.M.~O'Hara, J.H.~Huckans, S.L.~Rolston, W.D.~Phillips, and J.V.~Porto,
  Phys. Rev. Lett. \textbf{94}, 120403 (2005). 


\bibitem{Girardeau60-65} M. Girardeau, J. Math. Phys. \textbf{1}, 516
      (1960); M. Girardeau, Phys. Rev.  \textbf{139}, B500 (1965).

\bibitem{Lieb1961} E.~Lieb, T.~Shultz and D. Mattis, Ann. Phys. (NY)
  \textbf{16}, 406 (1961)

\bibitem{Schultz1963} T.D.~Schultz, J. Math. Phys. \textbf{4}, 666 (1963).

\bibitem{Efetov1975} K.B.~Efetov and A.I.~Larkin, Sov. Phys. JETP \textbf{42},
  390 (1976).

\bibitem{Vaidya1978} H.G.~Vaidya and C.A.~Tracy, Physica A
    \textbf{92}, 1 (1978).

\bibitem{McCoy1968-71} B.M.~McCoy, Phys. Rev. \textbf{127}, 1508
  (1962); E.~Barouch and B.M.~McCoy, Phys. Rev. A \textbf{3}, 786
  (1971); Phys. Rev. A \textbf{3}, 2137 (1971); B.M.~McCoy, E.~Barouch
  and D.~Abraham, Phys. Rev. A \textbf{4}, 2331 (1971).


\bibitem{Rigol2004} M.~Rigol and A.~Muramatsu, Phys. Rev. A \textbf{70},
  031603(R) (2004); Phys. Rev. A \textbf{72}, 013604 (2005).

\bibitem{KamenevMezard1999} A.~Kamenev and M.~Mezard, J.Phys. {\bf A 32},
4373 (1999), A.~Kamenev and M.~Mezard, Phys.Rev. B \textbf{60}, 3944
(1999).

\bibitem{GangardtKamenev2001} D.M.~Gangardt and A.~Kamenev, Nucl.~Phys.~B
\textbf{610}, 578 (2001); S.M. Nishigaki, D.M. Gangardt and
A. Kamenev, J.~Phys. A \textbf{36}, 3137 (2003).

\bibitem{Astrakharchik2006} G.E.~Astrakharchik, D.M.~Gangardt, Yu.E.~Lozovik,
  I.A.~Sorokin, cond-mat/0512470 .



\bibitem{Gangardt2004} D.M.~Gangardt, J. Phys. A, \textbf{ 37}, 9335 (2004). 

\bibitem{Edwards1975} S.F~Edwards and P.W.~Anderson, J. Phys. F
  \textbf{5}, 89 (1975).

\bibitem{Basor1994} E.L.~Basor and K.E.~Morrison, Linear
  Algebr. Appl. \textbf{202}, 129, (1994).

\bibitem{MehtaRandMatr} M. Mehta, {\it Random Matrices} (Academic
  Press, Boston, 1991), 2nd ed.

\bibitem{VaidyaTracy1979} H.G.~Vaidya and C.A.~Tracy,
  Phys. Rev. Lett. \textbf{42}, 3 (1979); J. Math. Phys. \textbf{20},
  2291 (1979).

\bibitem{Haldane1981} F.D.M.~Haldane, Phys. Rev. Lett. \textbf{47},
  1840 (1981).

\bibitem{Richard2003} S.~Richard, F.~Gerbier, J.H.~Thywissen, M.~Hugbart,
  P.~Bouyer, and A. Aspect, Phys.~Rev.~Lett. \textbf{91}, 010405 (2003).

\bibitem{SpinChainCorr} M.~Jimbo and T.~Miwa, \textit{Algerbraic analysis of
    solvable lattice models} (AMS, 1995); N.~Kitanine, J.M.~Maillet,
  N.A.~Slavnov, and V.~Terras in ``Solvable Lattice Models 2004'' 
  (edited by J.~Shirashi and Y.~Yamada) RIMS, Kyoto (2006).


 \bibitem{MehtaMatr} M. Mehta, {\it Theory of Matrices} (Academic
   Press, Boston, 1991), 2nd ed.


 \bibitem{GradshteynRyzhik} I.S.~Gradshteyn, I.M.~Ryzhik, \textit{Table
   of Integrals, Series, and Products} (Academic Press, Boston, 2000),
   6th ed.




\end{thebibliography}
\end{document}